\documentclass[%
 reprint,
 showpacs,preprintnumbers,
 amsmath,amssymb,
 aps,
]{revtex4-1}

\usepackage{graphicx}
\usepackage{dcolumn}
\usepackage{bm}

\begin{document}

\preprint{APS/123-QED}

\title{Self-Organized Cooperative Criticality in Coupled Complex Systems}

\author{Lei Liu}
\email{liulei@mail.iap.ac.cn}

\author{Fei Hu}%
\affiliation{%
 LAPC, Institute of Atmospheric Physics, Chinese Academy of
 Sciences, Beijing, China
}%

\date{\today}

\begin{abstract}
We show that the coupled complex systems can evolve into a new kind
of self-organized critical state where each subsystem is not
critical, however, they cooperate to be critical. This criticality
is different from the classical BTW criticality where the single
system itself evolves into a critical state. We also find that the
outflows can be accumulated in the coupled systems. This will lead
to the emergency of spatiotemporal intermittency in the critical
state.
\end{abstract}

\pacs{89.75.-k,05.40.-a,02.50-r}
\maketitle

Since proposed in 1980s, the self-organized criticality (SOC)
\cite{btw87,btw88,bak96} has been one of the most popular theory in
complex science and is widely used as a way of understanding
emergent complex behavior in physical
\cite{bt89,lh91,vh94,mrz99,bndp88}, biological
\cite{bs93,newman96,lnpi01} and even social systems
\cite{bcsw93,sw94,np95,rt98}. According to this theory, the
ubiquitous power law in nature can be interpreted as the hallmark of
critical state, a stable state that systems evolve into by
themselves. The SOC is usually illustrated by a simple cellular
automaton, the so-called BTW sandpile model
\cite{btw87,btw88,bak96}. The dynamics of this model can be imaged
as a transport phenomenon, where sand grains in a very high heap
(with a local height gradient exceeding the threshold) is unstable
and will tumble down the slope of this heap (against the local
height gradient). We note that this dynamics is not enough to
describe the real complex systems where a movement along the local
gradient is also possible. For example, in our daily life you may
have considered to migrate to a bigger city for a better job.
Migration to a bigger city is a movement along the population
gradient. Here, someone's final choice is determined by the
equilibrium between the population gradient and other gradients,
such as the job gradients, the education gradients and even shopping
gradients. In other words, the dynamics of migration is coupled with
other dynamics. In fact, similar coupling also exists in the natural
systems. A well-known example is the cross effect in the transport
arising in a mixture if both the concentrations and the temperature
are non-uniform over the system. The cross effect can be simply
illustrated in a isotropic binary mixture without viscosity when no
external forces are supposed to be present and the pressure is
uniform over the system \cite{gm84}:
\begin{eqnarray}
\mathbf{J}_{q}& = &-\lambda\nabla T-\alpha s_{T}\nabla C \label{eq:1}\\
\mathbf{J}_{d}& = &-\rho D\nabla C-\beta s_{T}\nabla T. \label{eq:2}
\end{eqnarray}
Symbols $\mathbf{J}_{q}$, $\mathbf{J}_{d}$, $T$, $C$ and $\rho$ are
the heat flux, the diffusion flow, the temperature, the
concentration of one of the components in the mixture and the
density of mixture respectively. The diffusion coefficient $D$ and
the heat conductivity $\lambda$ are positive and are related to the
normal heat conduction and diffusion respectively. The symbol
$s_{T}$ is called Soret coefficient which is related to the cross
effect in the transport and can be negative or positive. The
coefficients $\alpha\geq 0$ and $\beta \geq 0$. From
Eqs.~(\ref{eq:1}) and (\ref{eq:2}), one can see that the direction
of heat or diffusion flow is determined by the equilibrium between
the local temperature and the local concentration gradient. Two
transport processes are coupled together. From above discussions, a
question arises as to what will happen if the dynamics of two SOC
systems are coupled together? In this paper, we try to answer this
question by using a simple model.

The model we proposed here is composed of two BTW sandpiles. A BTW
sandpile is represented by a two-dimensional grid with a length of
$N$. At each square of the grid with coordinates $(x,y)$, we assign
a number $z(x,y)$, which represents the negative value of local
height gradient at that square. The coordinates $1\leq x\leq N$ and
$1\leq y\leq N$. Note that what $z$ represents in the sandpile model
is not important. The more important thing is its dynamics. In real
systems, $z$ would represent any interesting dynamical variable (see
examples in
Refs.~\cite{bt89,lh91,vh94,mrz99,bndp88,bs93,newman96,lnpi01,bcsw93,sw94,np95,rt98}).
Different from the dynamics of BTW sandpile, our model couple two
sandpiles in the way that {\it if and only if} both the negative
value of local height gradients of $z1$ and $z2$ of the two
sandpiles at the same positions ($x$,$y$) are greater than the
threshold $z_c$ (equaling $3$ in the following simulations), the
sand grains then tumble down,
\begin{equation}
z1,z2(x,y)=z1,z2(x,y)-4 \label{eq:3}
\end{equation}
and the tumbledown grains are transported to the nearest-neighbor
grids,
\begin{eqnarray}
z1,z2(x\pm 1,y)& = &z1,z2(x\pm 1,y)+1 \label{eq:4}\\
z1,z2(x,y\pm 1)& = &z1,z2(x,y\pm 1)+1 \label{eq:5}
\end{eqnarray}
This model is conservative except in the boundary, for example,
\begin{eqnarray*}
z1,z2(N,N)& = &z1,z2(N,N)-4 \\
z1,z2(N-1,N)& = &z1,z2(N-1,N)+1 \\
z1,z2(N,N-1)& = &z1,z2(N,N-1)+1.
\end{eqnarray*}
Two sandpiles are statistically symmetric and we just show the
statistics of one of them in the following analysis.

\begin{figure}[t]
\includegraphics[width=20pc]{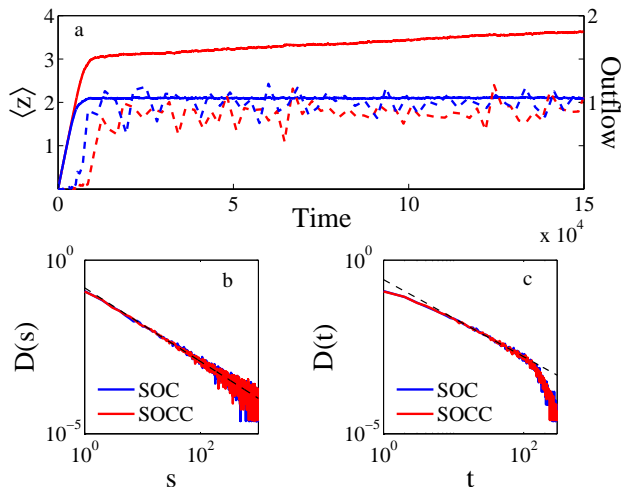}
\caption{\label{fig:1} Comparison between the BTW sandpile (SOC) and
the conservative coupled sandpiles (SOCC). (a) Time evolution of the
$\langle z\rangle$ averaged over the whole sandpile (line, red for
SOCC and blue for SOC) and the coarse grained outflow (dashed line,
red for SOCC and blue for SOC). (b) Distribution of avalanche sizes
$D(s)$ in critical states. The slope of the black dashed line is
about -1.06. (c) Distribution of lifetimes $D(t)$ in critical
states. The slope of the black dashed line is about -1.11.}
\end{figure}

{\it Self-Organized Cooperative Criticality (SOCC).} This is the
most important feature we found in the coupled sandpiles. Starting
from two sandpiles with the same length of $N$ (equaling $50$ in the
following simulations) and the zero local height gradients in the
whole sandpiles, the critical states are built by adding sand grains
randomly to each sandpile simultaneously according to the rule:
\begin{equation}
z1,z2(x,y)=z1,z2(x,y)+1, \label{eq:6}
\end{equation}
which is called the nonconservative perturbation in the BTW sandpile
model \cite{jensen}. After each perturbation, the two coupled
sandpiles are allowed to tumble down (if needed) according to
Eqs.~(\ref{eq:3}), (\ref{eq:4}) and (\ref{eq:5}). A local
perturbation would spread to the nearest-neighbor grids. These grids
would tumble down again and the perturbation is transported to the
next-nearest-neighbor grids and so on until the unstable grids (i.e.
$z1,z2(x,y)>z_c$) are not found in the sandpiles. Then another
perturbations are randomly added into each sandpile to trigger a new
possible avalanche.

As time elapses, the coupled sandpiles will evolve into a
self-organized critical state. A signal for the critical state is
the outflow which is defined as the total number of sand grains
flowing out of one sandpile after each perturbation. Both for the
BTW sandpile (blue broken line in Fig.~\ref{fig:1}a) and for the
coupled sandpiles (red broken line in Fig.~1a), the outflow will
fluctuate around a constant once the critical states are built.
There are other signals for the critical state, such as the
distributions of the avalanche size (also called the total energy
release) and the lifetime of avalanche
\cite{btw87,btw88,bak96,jensen}. The avalanche size $s$ is defined
as the total number of tumbles in one sandpile induced by each
perturbation and the lifetime of avalanche $t$ is defined as the
total number of simultaneous tumbles in one sandpile induced by each
perturbation. In the simulation, perturbations are not
simultaneously added into the two sandpiles and we will randomly
choose a sandpile for each perturbation. We count the avalanche size
$s$ and the lifetime $t$ induced by each perturbation and find that
in the critical states the distribution of avalanche sizes $D(s)$
and the distribution of lifetimes $D(t)$ follow the same power laws
both in the coupled sandpiles and the BTW sandpile
(Fig.~\ref{fig:1}b and \ref{fig:1}c).

Although the coupled sandpiles and the BTW sandpile have the same
distributions of avalanche sizes and lifetimes, their self-organized
critical states are very different in nature. For the BTW sandpile,
all $z(x,y)\leq z_c$ in the critical state. For the coupled
sandpiles, there are many ``supcritical'' grids with $z1>z_c$ or
$z2>z_c$ which mean that each separate sandpile in the coupled model
is not critical in view of the BTW sandpile. However, they are
indeed in the self-organized critical state not separately but
cooperatively. Cooperative phenomena are common in the complex
systems. Many species, such as bees and ants, can behave
cooperatively in a self-organized way (see Ref.~\cite{fisher09} and
references therein). Our model join the two important notions of the
cooperation and the SOC together in a simple manner. Thus, we call
the new critical mode the {\it self-organized cooperative
criticality (SOCC)}.

\begin{figure}[b]
\includegraphics[width=18pc]{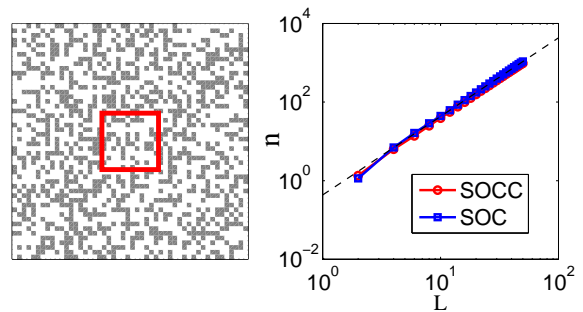}
\caption{\label{fig:2} ({\it left}) The snapshot of the SOCC state
in one sandpile. The minimally stable states with ``$z1>z_c$ and
$z2=z_c$'' or ``$z2>z_c$ and $z1=z_c$'' are denoted by grey grids.
({\it right}) The relation between the number of minimally stable
states $n$ in the chosen rectangle (as the red rectangle in the left
plot) and the length of this rectangle $L$, averaged over 5000
samples. The slope of the dashed line is 2.}
\end{figure}

The left plot of Fig.~\ref{fig:2} shows a snapshot of the structure
of minimally stable state with ``$z1=z_c$ and $z2>z_c$'' or
``$z1>z_c$ and $z2=z_c$'' in one sandpile. It likes a fractal
structure and its dimension $D$ can be measured as follows
\cite{feder}. We lay a square in the center of whole sandpile and
count the number of minimally stable states $n$ in this square.
Then, a power-law relation between the length of the square $L$ and
$n$ can be set up. That is the length-number relation: $n\sim
L^{D}$, where $D$ is called the cluster fractal dimension. Measured
result shows that $D=\log n/\log L\approx 2$ (the right plot of
Fig.~\ref{fig:2}), which means that the structure of minimally
stable state in the coupled sandpiles is a fat fractal \cite{ott}.
More interesting, the structure of minimally stable state (with
$z=z_c$) in the BTW sandpile is also a fat fractal with the same
cluster fractal dimension $D\approx 2$. Thus, the geometry
structures of minimally stable state are statistically similar both
in the SOC and SOCC sandpiles. That is possibly why we obtain the
similar distributions of avalanche size and lifetimes in the two
models.

\begin{figure}[t]
\includegraphics[width=20pc]{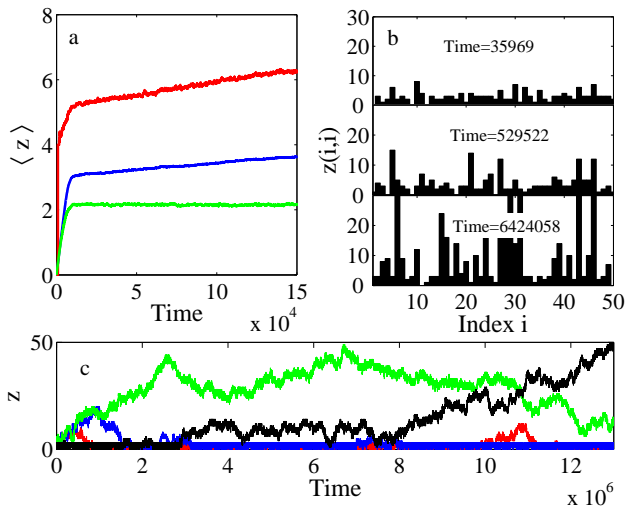}
\caption{\label{fig:3} Accumulative effect and spatiotemporal
intermittency. (a) Time evolution of the $\langle z\rangle$ averaged
over the grids with $z>z_c$ (red), $z\leq z_c$ (green) and all $z$
(blue). (b) The values of $z$ along the diagonal direction of one
sandpile at different times. (c) Time evolution of $z$ at four
randomly chosen grids with coordinates $(5,5)$ (blue), $(25,25)$
(red), $(7,30)$(green) and $(24,6)$(black).}
\end{figure}

\begin{figure}[b]
\includegraphics[width=20pc]{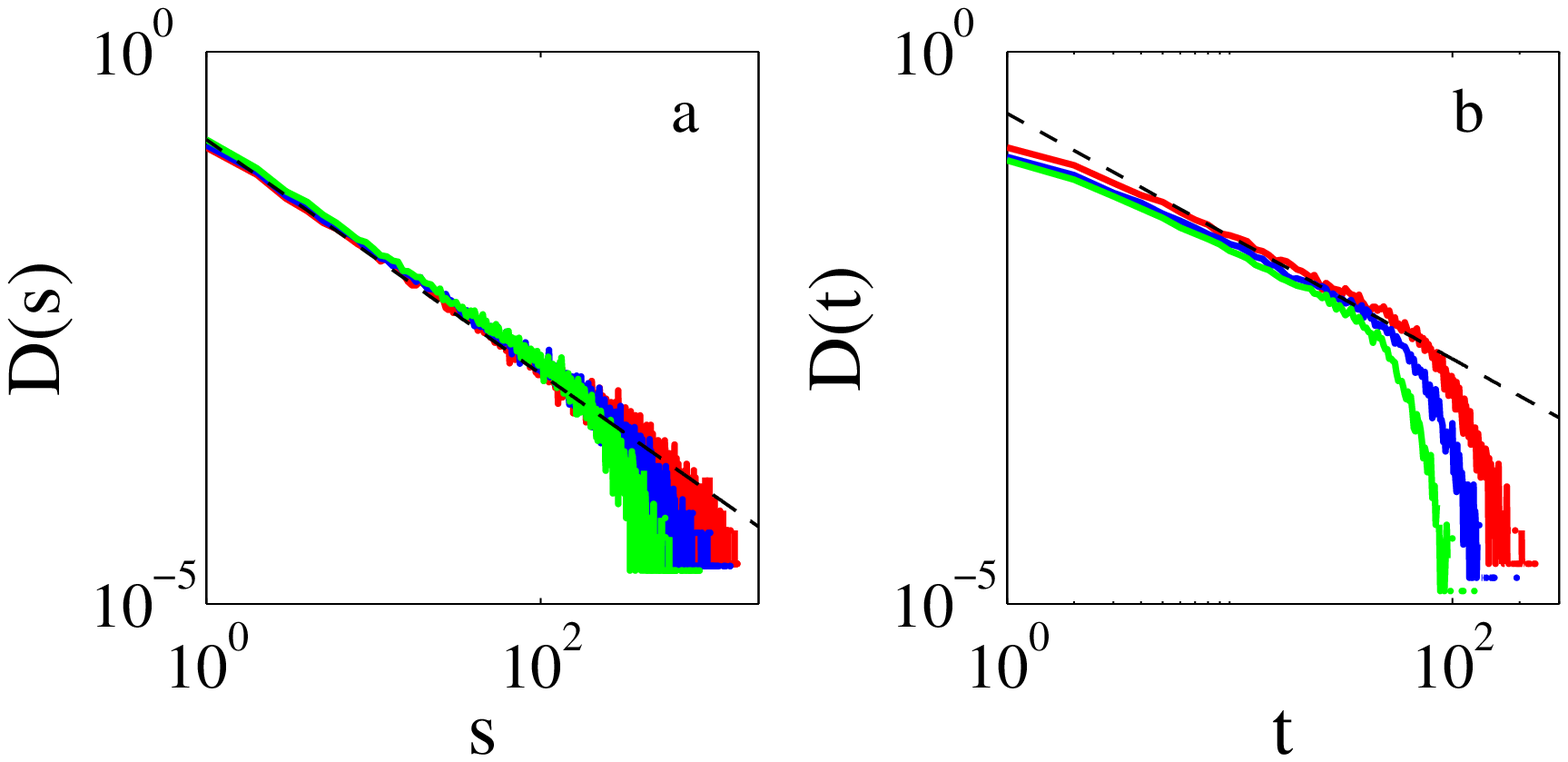}\\
\includegraphics[width=20pc]{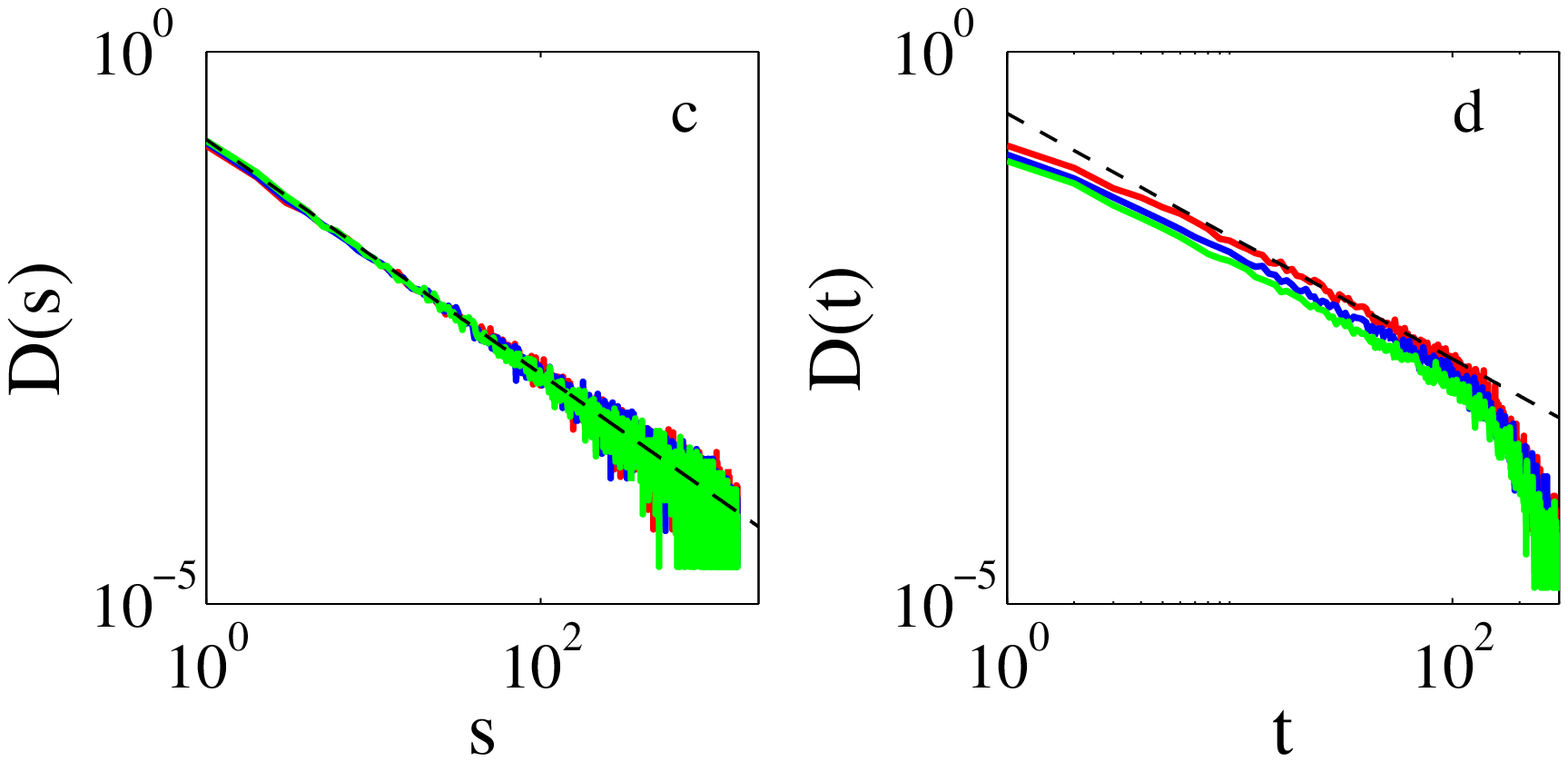}
\caption{\label{fig:4} Distributions of the avalanche sizes and the
lifetimes in the SOCC state of the spoiled couple sandpiles. In (a)
and (b), the model is spoiled by randomly removing the connections
between nearest-neighbor grids in each sandpile. The lines with
different colors correspond to different numbers of removed
connections in each sandpile (red for 250, blue for 500 and green
for 1000). In (c) and (d), the model is spoiled by randomly removing
the correlations between the two coupled sandpiles. The meaning of
different colors is the same as in (a) and (b). The slopes of black
dashed lines in all above figure are the same in Fig.~\ref{fig:1}b
and \ref{fig:1}c.}
\end{figure}

{\it Accumulative Effect and Spatiotemporal Intermittency}. We note
that the outflow in the SOC state fluctuates around a value of 1
(the blue broken line in Fig.~\ref{fig:1}a). This means that the
inflow and the outflow in the SOC state is in statistical
equilibrium and the average of $z(x,y)$ will also fluctuate around a
constant (the blue lines in Fig.~\ref{fig:1}a). However, in the SOCC
state the outflow fluctuates around a constant smaller than 1 (the
red broken line in Fig.~\ref{fig:1}a). This means that the inflow is
greater than the outflow and the adding grains will be accumulated
in the sandpile in a constant rate (the red line in
Fig.~\ref{fig:1}a).

In Fig.~\ref{fig:3}a, we analyze the averages of $z$ with $z>z_c$
and $z\leq z_c$ respectively and find that the adding grains are
mainly accumulated to the ``supcritical'' grids with $z<z_c$. To see
the accumulative process intuitively, we observe the evolution of
the distribution of $z$ in the diagonal direction in one sandpile
(Fig.~\ref{fig:3}b). One can see that as time elapses the grids with
$z>z_c$ accumulate more and more adding grains. At the same time, a
more and more intermittent distribution of $z$ appears in space
(Fig.~\ref{fig:3}b). The intermittency will also appear in the time
series of $z$ at a fixed grid (Fig.~\ref{fig:3}c). In all, we
observe the spatiotemporal intermittency of local height gradients
in the coupled sandpiles. Many complex systems has the
spatiotemporal intermittency, such as the fluid turbulence
\cite{frisch,mbp10,bm10}, the earthquakes \cite{bak96,bg07} and the
traffic jam \cite{np95}. The accumulative effect of the SOCC
sandpiles provides another possible mechanism for the emergency of
spatiotemporal intermittency.

{\it Robustness.} To test the robustness of the criticality in the
coupled model, we randomly remove the connections between
nearest-neighbor grids in each sandplie. When the connection between
nearest-neighbor grids, say $z1(x,y)$ and $z1(x+1,y)$, is removed,
the rules of Eqs.~(\ref{eq:3}), (\ref{eq:4}) and (\ref{eq:5}) are
modified by
\begin{eqnarray*}
z1(x,y) & = &z1(x,y)-3\\
z1(x+1,y)& = &z1(x+1,y) \\
z1(x-1,y)& = &z1(x-1,y)+1 \\
z1(x,y\pm 1)& = &z1(x,y\pm 1)+1.
\end{eqnarray*}
In the simulation, the numbers of removed connections for the two
sandpiles are the same. Figures \ref{fig:4}a and \ref{fig:4}b show
the distributions of $D(s)$ and $D(t)$ with different numbers of
removed connections. We find that these distributions also have the
power laws and no changes have been found in the exponents. However,
the statistics on the very large avalanches deviate from the power
laws. With the increase of removal, the deviation becomes large. The
connectivity of the sandpile is decreased by the removal. Thus, the
large avalanches occur with a less frequency than in the unspoiled
sandpile.

We also remove the correlations between the two sandpiles at the
same coordinations. At these grids without correlations, if $z1>z_c$
or $z2>z_c$ the sand grains will tumble down. Simulations show that
almost nothing is changed in the distributions of $D(s)$ and $D(t)$
even the proportion of the removed correlations is increased to 20\%
(Figs.~\ref{fig:4}c and \ref{fig:4}d ).

\begin{figure}[t]
\includegraphics[width=20pc]{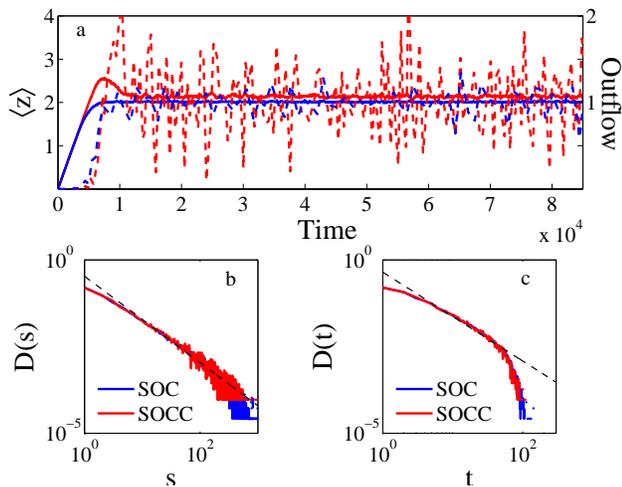}
\caption{\label{fig:5} Comparison between the BTW sandpile (SOC) and
the non-conservative coupled sandpiles (SOCC). The meanings of
symbols are the same as in Fig.~\ref{fig:1}. The slopes of black
dashed lines in (b) and (c) are about -1.25 and -1.28 respectively.
}
\end{figure}

{\it Non-conservative Coupled Sandpiles.} The average of $z$ in the
conservative coupled sandpiles will increase infinitely if we
continually add sand grains into the sandpiles. It seems unphysical.
In real systems, the dissipation always exists. In
Ref.~\cite{feders}, authors proposed a non-conservative BTW sandpile
to simulate the effect of dissipation. In this model, when tumbles
happen at $(x,y)$, no matter how large $z(x,y)$ is, it will become
zero. Following this sprit, we modify the rules of (\ref{eq:3}) by
$z1,z2(x,y)=0$. Other rules like Eqs.~(\ref{eq:4}) and (\ref{eq:5})
are not changed.

The non-conservative coupled sandpiles will also evolve into the
SOCC states. Different from the conservative sandpiles, the inflow
and outflow in this SOCC state is in statistical equilibrium. It
means that the accumulation effect is statistically offset by the
dissipation. We stress ``statistically'' here because the local
grids with $z>z_c$ will also accumulate sand grains if tumbles have
not happened in these grids. We measure the distributions of $D(s)$
and $D(t)$ in the critical states and find that they also have the
power laws with the same exponents as in the conservative BTW
sandpile (Figs.~\ref{fig:5}b and \ref{fig:5}c).

In conclusion, we propose a model by coupling two BTW sandpiles in a
simple way. This model can conjoin many notions popular in the
complex science, including the SOC, the cooperation, and the
spatiotemporal intermittency. We believe that our model could be
further extended to understand the real systems which might be
composed of many coupled SOC systems.

\begin{acknowledgments}
This work was supported by the National Nature Science Foundation of
China under Grant No. 41105005 and the Strategic Project of Science
and Technology of Chinese Academy of Sciences under Grant No.
XDA05040301.
\end{acknowledgments}

\bibliography{liuhu_manuscript}

\providecommand{\noopsort}[1]{}\providecommand{\singleletter}[1]{#1}%
\begin{thebibliography}{25}%
\makeatletter
\providecommand \@ifxundefined [1]{%
 \@ifx{#1\undefined}
}%
\providecommand \@ifnum [1]{%
 \ifnum #1\expandafter \@firstoftwo
 \else \expandafter \@secondoftwo
 \fi
}%
\providecommand \@ifx [1]{%
 \ifx #1\expandafter \@firstoftwo
 \else \expandafter \@secondoftwo
 \fi
}%
\providecommand \natexlab [1]{#1}%
\providecommand \enquote  [1]{``#1''}%
\providecommand \bibnamefont  [1]{#1}%
\providecommand \bibfnamefont [1]{#1}%
\providecommand \citenamefont [1]{#1}%
\providecommand \href@noop [0]{\@secondoftwo}%
\providecommand \href [0]{\begingroup \@sanitize@url \@href}%
\providecommand \@href[1]{\@@startlink{#1}\@@href}%
\providecommand \@@href[1]{\endgroup#1\@@endlink}%
\providecommand \@sanitize@url [0]{\catcode `\\12\catcode `\$12\catcode
  `\&12\catcode `\#12\catcode `\^12\catcode `\_12\catcode `\%12\relax}%
\providecommand \@@startlink[1]{}%
\providecommand \@@endlink[0]{}%
\providecommand \url  [0]{\begingroup\@sanitize@url \@url }%
\providecommand \@url [1]{\endgroup\@href {#1}{\urlprefix }}%
\providecommand \urlprefix  [0]{URL }%
\providecommand \Eprint [0]{\href }%
\providecommand \doibase [0]{http://dx.doi.org/}%
\providecommand \selectlanguage [0]{\@gobble}%
\providecommand \bibinfo  [0]{\@secondoftwo}%
\providecommand \bibfield  [0]{\@secondoftwo}%
\providecommand \translation [1]{[#1]}%
\providecommand \BibitemOpen [0]{}%
\providecommand \bibitemStop [0]{}%
\providecommand \bibitemNoStop [0]{.\EOS\space}%
\providecommand \EOS [0]{\spacefactor3000\relax}%
\providecommand \BibitemShut  [1]{\csname bibitem#1\endcsname}%
\let\auto@bib@innerbib\@empty
\bibitem [{\citenamefont {Bak}\ \emph {et~al.}(1987)\citenamefont {Bak},
  \citenamefont {Tang},\ and\ \citenamefont {Wiesenfeld}}]{btw87}%
  \BibitemOpen
  \bibfield  {author} {\bibinfo {author} {\bibfnamefont {P.}~\bibnamefont
  {Bak}}, \bibinfo {author} {\bibfnamefont {C.}~\bibnamefont {Tang}}, \ and\
  \bibinfo {author} {\bibfnamefont {K.}~\bibnamefont {Wiesenfeld}},\
  }\href@noop {} {\bibfield  {journal} {\bibinfo  {journal} {Phys.\ Rev.\
  Lett.}\ }\textbf {\bibinfo {volume} {59}},\ \bibinfo {pages} {381} (\bibinfo
  {year} {1987})}\BibitemShut {NoStop}%
\bibitem [{\citenamefont {Bak}\ \emph {et~al.}(1988)\citenamefont {Bak},
  \citenamefont {Tang},\ and\ \citenamefont {Wiesenfeld}}]{btw88}%
  \BibitemOpen
  \bibfield  {author} {\bibinfo {author} {\bibfnamefont {P.}~\bibnamefont
  {Bak}}, \bibinfo {author} {\bibfnamefont {C.}~\bibnamefont {Tang}}, \ and\
  \bibinfo {author} {\bibfnamefont {K.}~\bibnamefont {Wiesenfeld}},\
  }\href@noop {} {\bibfield  {journal} {\bibinfo  {journal} {Phys.\ Rev.\ A}\
  }\textbf {\bibinfo {volume} {38}},\ \bibinfo {pages} {364} (\bibinfo {year}
  {1988})}\BibitemShut {NoStop}%
\bibitem [{\citenamefont {Bak}(1996)}]{bak96}%
  \BibitemOpen
  \bibfield  {author} {\bibinfo {author} {\bibfnamefont {P.}~\bibnamefont
  {Bak}},\ }\href@noop {} {\emph {\bibinfo {title} {How Nature Works}}}\
  (\bibinfo  {publisher} {Springer-Verlag New York,Inc.},\ \bibinfo {year}
  {1996})\BibitemShut {NoStop}%
\bibitem [{\citenamefont {Bak}\ and\ \citenamefont {Tang}(1989)}]{bt89}%
  \BibitemOpen
  \bibfield  {author} {\bibinfo {author} {\bibfnamefont {P.}~\bibnamefont
  {Bak}}\ and\ \bibinfo {author} {\bibfnamefont {C.}~\bibnamefont {Tang}},\
  }\href@noop {} {\bibfield  {journal} {\bibinfo  {journal} {J.\ Geophys.\
  Res.}\ }\textbf {\bibinfo {volume} {B94}},\ \bibinfo {pages} {15635}
  (\bibinfo {year} {1989})}\BibitemShut {NoStop}%
\bibitem [{\citenamefont {Lu}\ and\ \citenamefont {Hamilton}(1991)}]{lh91}%
  \BibitemOpen
  \bibfield  {author} {\bibinfo {author} {\bibfnamefont {E.~T.}\ \bibnamefont
  {Lu}}\ and\ \bibinfo {author} {\bibfnamefont {R.~J.}\ \bibnamefont
  {Hamilton}},\ }\href@noop {} {\bibfield  {journal} {\bibinfo  {journal}
  {Astrophys.\ J.}\ }\textbf {\bibinfo {volume} {380}},\ \bibinfo {pages} {L89}
  (\bibinfo {year} {1991})}\BibitemShut {NoStop}%
\bibitem [{\citenamefont {Vattay}\ and\ \citenamefont {Harnos}(1994)}]{vh94}%
  \BibitemOpen
  \bibfield  {author} {\bibinfo {author} {\bibfnamefont {G.}~\bibnamefont
  {Vattay}}\ and\ \bibinfo {author} {\bibfnamefont {A.}~\bibnamefont
  {Harnos}},\ }\href@noop {} {\bibfield  {journal} {\bibinfo  {journal} {Phys.\
  Rev.\ Lett.}\ }\textbf {\bibinfo {volume} {73}},\ \bibinfo {pages} {768}
  (\bibinfo {year} {1994})}\BibitemShut {NoStop}%
\bibitem [{\citenamefont {Meng}\ \emph {et~al.}(1999)\citenamefont {Meng},
  \citenamefont {Rittel},\ and\ \citenamefont {Zhang}}]{mrz99}%
  \BibitemOpen
  \bibfield  {author} {\bibinfo {author} {\bibfnamefont {T.-C.}\ \bibnamefont
  {Meng}}, \bibinfo {author} {\bibfnamefont {R.}~\bibnamefont {Rittel}}, \ and\
  \bibinfo {author} {\bibfnamefont {Y.}~\bibnamefont {Zhang}},\ }\href@noop {}
  {\bibfield  {journal} {\bibinfo  {journal} {Phys.\ Rev.\ Lett.}\ }\textbf
  {\bibinfo {volume} {82}},\ \bibinfo {pages} {2044} (\bibinfo {year}
  {1999})}\BibitemShut {NoStop}%
\bibitem [{\citenamefont {Carreras}\ \emph {et~al.}(2004)\citenamefont
  {Carreras}, \citenamefont {Newman}, \citenamefont {Dobson},\ and\
  \citenamefont {Poole}}]{bndp88}%
  \BibitemOpen
  \bibfield  {author} {\bibinfo {author} {\bibfnamefont {B.~A.}\ \bibnamefont
  {Carreras}}, \bibinfo {author} {\bibfnamefont {D.~E.}\ \bibnamefont
  {Newman}}, \bibinfo {author} {\bibfnamefont {I.}~\bibnamefont {Dobson}}, \
  and\ \bibinfo {author} {\bibfnamefont {A.~B.}\ \bibnamefont {Poole}},\
  }\href@noop {} {\bibfield  {journal} {\bibinfo  {journal} {IEEE Trans.\
  Circuits Syst., I: Fundam.\ Theory Appl.}\ }\textbf {\bibinfo {volume}
  {51}},\ \bibinfo {pages} {1733} (\bibinfo {year} {2004})}\BibitemShut
  {NoStop}%
\bibitem [{\citenamefont {Bak}\ and\ \citenamefont {Sneppen}(1993)}]{bs93}%
  \BibitemOpen
  \bibfield  {author} {\bibinfo {author} {\bibfnamefont {P.}~\bibnamefont
  {Bak}}\ and\ \bibinfo {author} {\bibfnamefont {K.}~\bibnamefont {Sneppen}},\
  }\href@noop {} {\bibfield  {journal} {\bibinfo  {journal} {Phys.\ Rev.\
  Lett.}\ }\textbf {\bibinfo {volume} {71}},\ \bibinfo {pages} {4083} (\bibinfo
  {year} {1993})}\BibitemShut {NoStop}%
\bibitem [{\citenamefont {Newman}(1996)}]{newman96}%
  \BibitemOpen
  \bibfield  {author} {\bibinfo {author} {\bibfnamefont {M.~E.~J.}\
  \bibnamefont {Newman}},\ }\href@noop {} {\bibfield  {journal} {\bibinfo
  {journal} {Proc.\ R.\ Soc.\ Lond.\ B}\ }\textbf {\bibinfo {volume} {263}},\
  \bibinfo {pages} {1605} (\bibinfo {year} {1996})}\BibitemShut {NoStop}%
\bibitem [{\citenamefont {Linkenkaer-Hansen}\ \emph {et~al.}(2001)\citenamefont
  {Linkenkaer-Hansen}, \citenamefont {Nikouline}, \citenamefont {Palva},\ and\
  \citenamefont {Ilmoniemi}}]{lnpi01}%
  \BibitemOpen
  \bibfield  {author} {\bibinfo {author} {\bibfnamefont {K.}~\bibnamefont
  {Linkenkaer-Hansen}}, \bibinfo {author} {\bibfnamefont {V.~V.}\ \bibnamefont
  {Nikouline}}, \bibinfo {author} {\bibfnamefont {J.~M.}\ \bibnamefont
  {Palva}}, \ and\ \bibinfo {author} {\bibfnamefont {R.~J.}\ \bibnamefont
  {Ilmoniemi}},\ }\href@noop {} {\bibfield  {journal} {\bibinfo  {journal} {J.\
  Neurosci.}\ }\textbf {\bibinfo {volume} {21}},\ \bibinfo {pages} {1370}
  (\bibinfo {year} {2001})}\BibitemShut {NoStop}%
\bibitem [{\citenamefont {Bak}\ \emph {et~al.}(1993)\citenamefont {Bak},
  \citenamefont {Chen}, \citenamefont {Scheinkman},\ and\ \citenamefont
  {Woodford}}]{bcsw93}%
  \BibitemOpen
  \bibfield  {author} {\bibinfo {author} {\bibfnamefont {P.}~\bibnamefont
  {Bak}}, \bibinfo {author} {\bibfnamefont {K.}~\bibnamefont {Chen}}, \bibinfo
  {author} {\bibfnamefont {J.}~\bibnamefont {Scheinkman}}, \ and\ \bibinfo
  {author} {\bibfnamefont {M.}~\bibnamefont {Woodford}},\ }\href@noop {}
  {\bibfield  {journal} {\bibinfo  {journal} {Ric.\ Econ.}\ }\textbf {\bibinfo
  {volume} {47}},\ \bibinfo {pages} {3} (\bibinfo {year} {1993})}\BibitemShut
  {NoStop}%
\bibitem [{\citenamefont {Scheinkman}\ and\ \citenamefont
  {Woodford}(1994)}]{sw94}%
  \BibitemOpen
  \bibfield  {author} {\bibinfo {author} {\bibfnamefont {J.~A.}\ \bibnamefont
  {Scheinkman}}\ and\ \bibinfo {author} {\bibfnamefont {M.}~\bibnamefont
  {Woodford}},\ }\href@noop {} {\bibfield  {journal} {\bibinfo  {journal} {Am.\
  Econ.\ Rev.}\ }\textbf {\bibinfo {volume} {84}},\ \bibinfo {pages} {417}
  (\bibinfo {year} {1994})}\BibitemShut {NoStop}%
\bibitem [{\citenamefont {Nagel}\ and\ \citenamefont {Paczuski}(1995)}]{np95}%
  \BibitemOpen
  \bibfield  {author} {\bibinfo {author} {\bibfnamefont {K.}~\bibnamefont
  {Nagel}}\ and\ \bibinfo {author} {\bibfnamefont {M.}~\bibnamefont
  {Paczuski}},\ }\href@noop {} {\bibfield  {journal} {\bibinfo  {journal}
  {Phys.\ Rev.\ E}\ }\textbf {\bibinfo {volume} {51}},\ \bibinfo {pages} {2909}
  (\bibinfo {year} {1995})}\BibitemShut {NoStop}%
\bibitem [{\citenamefont {Roberts}\ and\ \citenamefont
  {Turcotte}(1998)}]{rt98}%
  \BibitemOpen
  \bibfield  {author} {\bibinfo {author} {\bibfnamefont {D.~C.}\ \bibnamefont
  {Roberts}}\ and\ \bibinfo {author} {\bibfnamefont {D.~L.}\ \bibnamefont
  {Turcotte}},\ }\href@noop {} {\bibfield  {journal} {\bibinfo  {journal}
  {Fractals}\ }\textbf {\bibinfo {volume} {6}},\ \bibinfo {pages} {351}
  (\bibinfo {year} {1998})}\BibitemShut {NoStop}%
\bibitem [{\citenamefont {Groot}\ and\ \citenamefont {Mazur}(1984)}]{gm84}%
  \BibitemOpen
  \bibfield  {author} {\bibinfo {author} {\bibfnamefont {S.~R.~D.}\
  \bibnamefont {Groot}}\ and\ \bibinfo {author} {\bibfnamefont
  {P.}~\bibnamefont {Mazur}},\ }\href@noop {} {\emph {\bibinfo {title}
  {Non-equilibrium Thermodynamics}}}\ (\bibinfo  {publisher} {Dover
  Publications, Inc.},\ \bibinfo {year} {1984})\BibitemShut {NoStop}%
\bibitem [{\citenamefont {Jensen}(1998)}]{jensen}%
  \BibitemOpen
  \bibfield  {author} {\bibinfo {author} {\bibfnamefont {H.~J.}\ \bibnamefont
  {Jensen}},\ }\href@noop {} {\emph {\bibinfo {title} {Self-Organized
  Criticality: Emergent Complex Behavior in Physical and Biological Systems}}}\
  (\bibinfo  {publisher} {Cambridge University Press},\ \bibinfo {year}
  {1998})\BibitemShut {NoStop}%
\bibitem [{\citenamefont {Fisher}(2009)}]{fisher09}%
  \BibitemOpen
  \bibfield  {author} {\bibinfo {author} {\bibfnamefont {L.}~\bibnamefont
  {Fisher}},\ }\href@noop {} {\emph {\bibinfo {title} {The Perfect Swarm: The
  Science of Complexity in Everyday Life}}}\ (\bibinfo  {publisher} {Basic
  Books},\ \bibinfo {year} {2009})\BibitemShut {NoStop}%
\bibitem [{\citenamefont {Feder}(1988)}]{feder}%
  \BibitemOpen
  \bibfield  {author} {\bibinfo {author} {\bibfnamefont {J.}~\bibnamefont
  {Feder}},\ }\href@noop {} {\emph {\bibinfo {title} {Fractals}}}\ (\bibinfo
  {publisher} {Plenum},\ \bibinfo {year} {1988})\BibitemShut {NoStop}%
\bibitem [{\citenamefont {Ott}(1993)}]{ott}%
  \BibitemOpen
  \bibfield  {author} {\bibinfo {author} {\bibfnamefont {E.}~\bibnamefont
  {Ott}},\ }\href@noop {} {\emph {\bibinfo {title} {Chaos in Dynamical
  Systems}}}\ (\bibinfo  {publisher} {Cambridge University Press},\ \bibinfo
  {year} {1993})\BibitemShut {NoStop}%
\bibitem [{\citenamefont {Frisch}(1995)}]{frisch}%
  \BibitemOpen
  \bibfield  {author} {\bibinfo {author} {\bibfnamefont {U.}~\bibnamefont
  {Frisch}},\ }\href@noop {} {\emph {\bibinfo {title} {Turbulence: The Legacy
  of A.N. Kolmogorov}}}\ (\bibinfo  {publisher} {Cambridge University Press},\
  \bibinfo {year} {1995})\BibitemShut {NoStop}%
\bibitem [{\citenamefont {Muzy}\ \emph {et~al.}(2010)\citenamefont {Muzy},
  \citenamefont {Ba{\"{i}}le},\ and\ \citenamefont {Poggi}}]{mbp10}%
  \BibitemOpen
  \bibfield  {author} {\bibinfo {author} {\bibfnamefont {J.~F.}\ \bibnamefont
  {Muzy}}, \bibinfo {author} {\bibfnamefont {R.}~\bibnamefont {Ba{\"{i}}le}}, \
  and\ \bibinfo {author} {\bibfnamefont {P.}~\bibnamefont {Poggi}},\
  }\href@noop {} {\bibfield  {journal} {\bibinfo  {journal} {Phys.\ Rev.\ E}\
  }\textbf {\bibinfo {volume} {81}},\ \bibinfo {pages} {056308} (\bibinfo
  {year} {2010})}\BibitemShut {NoStop}%
\bibitem [{\citenamefont {Ba{\"{i}}le}\ and\ \citenamefont
  {Muzy}(2010)}]{bm10}%
  \BibitemOpen
  \bibfield  {author} {\bibinfo {author} {\bibfnamefont {R.}~\bibnamefont
  {Ba{\"{i}}le}}\ and\ \bibinfo {author} {\bibfnamefont {J.~F.}\ \bibnamefont
  {Muzy}},\ }\href@noop {} {\bibfield  {journal} {\bibinfo  {journal} {Phys.\
  Rev.\ Lett.}\ }\textbf {\bibinfo {volume} {105}},\ \bibinfo {pages} {254501}
  (\bibinfo {year} {2010})}\BibitemShut {NoStop}%
\bibitem [{\citenamefont {Bottiglieri}\ and\ \citenamefont
  {Godano}(2007)}]{bg07}%
  \BibitemOpen
  \bibfield  {author} {\bibinfo {author} {\bibfnamefont {M.}~\bibnamefont
  {Bottiglieri}}\ and\ \bibinfo {author} {\bibfnamefont {C.}~\bibnamefont
  {Godano}},\ }\href@noop {} {\bibfield  {journal} {\bibinfo  {journal} {Phys.\
  Rev.\ E}\ }\textbf {\bibinfo {volume} {75}},\ \bibinfo {pages} {026101}
  (\bibinfo {year} {2007})}\BibitemShut {NoStop}%
\bibitem [{\citenamefont {Feder}\ and\ \citenamefont {Feder}(1991)}]{feders}%
  \BibitemOpen
  \bibfield  {author} {\bibinfo {author} {\bibfnamefont {H.~J.~S.}\
  \bibnamefont {Feder}}\ and\ \bibinfo {author} {\bibfnamefont
  {J.}~\bibnamefont {Feder}},\ }\href@noop {} {\bibfield  {journal} {\bibinfo
  {journal} {Phys.\ Rev.\ Lett.}\ }\textbf {\bibinfo {volume} {66}},\ \bibinfo
  {pages} {2669} (\bibinfo {year} {1991})}\BibitemShut {NoStop}%
\end{thebibliography}%

\end{document}